# Near-Unity Spin Hall Ratio in Ni$_x$Cu$_{1-x}$ Alloys

Mark W. Keller[1*], Katy S. Gerace[1†], Monika Arora[1], Erna Krisztina Delczeg-Czirjak[2], Justin M. Shaw[1], and T. J. Silva[1]

[1]National Institute of Standards and Technology, Boulder, CO 80305

[2]Department of Physics and Astronomy, Uppsala University, Box 516, SE-751 20 Uppsala, Sweden

We report a large spin Hall effect in the 3$d$ transition metal alloy Ni$_x$Cu$_{1-x}$ for $x \in \{0.3, 0.75\}$, detected via the ferromagnetic resonance of a Permalloy (Py = Ni$_{80}$Fe$_{20}$) film deposited in a bilayer with the alloy. A thickness series at $x = 0.6$, for which the alloy is paramagnetic at room temperature, allows us to determine the spin Hall ratio $\theta_{SH} \approx 1$, spin diffusion length $\lambda_s$, spin mixing conductance $G_{\uparrow\downarrow}$, and damping due to spin memory loss $\alpha_{SML}$. We compare our results with similar experiments on Py/Pt bilayers measured using the same method. Ab initio band structure calculations with disorder and spin-orbit coupling suggest an intrinsic spin Hall effect in Ni$_x$Cu$_{1-x}$ alloys, although the experiments here cannot distinguish between extrinsic and intrinsic mechanisms.

## I. INTRODUCTION

Electrical control of the direction of the magnetization $M$ in a thin ferromagnetic (FM) layer in contact with nonmagnetic (NM) layers is the basis for a number of magnetic random-access memory (MRAM) devices. The Oersted field from current in a write line can be used to switch $M$, as in the Toggle-MRAM available commercially [1] since 2004, but Joule heating and the long-range nature of the Oersted field pose severe problems for scalability. A spin-polarized charge current flowing into a metallic FM layer can switch $M$ via the spin torque (ST) effect, as in the ST-MRAM available [2] for niche applications since 2014, but lower switching energy, higher speed, and better endurance are needed for ST-MRAM to be used more widely. Pure spin current (without an associated charge current) can be generated by spin pumping from a FM into an adjacent NM, or by charge-to-spin conversion at a FM/NM interface via the spin Hall effect (SHE) or the Rashba-Edelstein effect (REE). Since spin-orbit coupling (SOC) is essential for these last two effects, they are known as spin-orbit torque (SOT) effects, and they have been proposed as a path to lower power and higher speed in future MRAM [3].

---

[*] Corresponding author; mark.keller@nist.gov
[†] Currently at Pennsylvania State University

The number of NM materials demonstrated to have SOT effects strong enough to enable efficient MRAM or other applications is fairly small. A rough criterion for "strong enough" is that the spin Hall ratio $\theta_{SH}$, defined as the ratio of spin current to charge current that flows in response to an applied electric field, be at least 10 %. The search for strong SOT materials has largely focused on metals with large atomic number $Z$ because of their strong SOC. Indeed, among pure elements only the 5$d$ transition metals Au, Hf, Pt, Ta, and W have shown $\theta_{SH} \geq 0.1$ [4, 5]. However, reports of appreciable SHE in much lighter 3$d$ transition metals [6, 7], including those with magnetic ordering [8, 9], show that SOT effects depend on more than SOC alone. In the case of the intrinsic SHE mechanism, it is known that details of the electronic band structure, in particular the Berry curvature near specific points on the Fermi surface, are essential to the effect [4, 10].

Our investigation of $Ni_xCu_{1-x}$ was motivated by an intriguing similarity between the Fermi surfaces of Pt and paramagnetic Ni that suggests a similar intrinsic SHE in both materials. Tight-binding calculations [11] for Pt and Ni (fcc, paramagnetic) show that both metals have a closed electron surface centered at the $\Gamma$ point with a narrow neck that reaches the Brillioun zone at the L points, as well as hole pockets localized near the X points. In ab initio calculations for Pt [10], a large intrinsic spin Hall conductivity was attributed to gaps opened by SOC at both of these high symmetry points. A similar but smaller lifting of degeneracy at these points is expected in paramagnetic Ni and is observed in the calculations reported here. Producing paramagnetic Ni by heating it above the Curie temperature, $T_c$ = 630 K, is inconvenient for experiments, but alloying with Cu decreases $T_c$ rapidly [12]. Our band structure calculations for $Ni_xCu_{1-x}$ show that the features near the L and X points are recognizable despite broadening due to alloy disorder, suggesting that there is reason to expect a significant intrinsic SHE in the alloy.

The experimental work reported here involves measuring the ferromagnetic resonance (FMR) of a Permalloy (Py = $Ni_{80}Fe_{20}$) film deposited in a bilayer with $Ni_xCu_{1-x}$ for $x \in \{0.3, 0.75\}$. The contributions of SOT effects can be extracted from the FMR signal using methods we have described previously [13]. Measurements on an alloy thickness series at $x$ = 0.6, for which the alloy is paramagnetic at room temperature, allow us to determine spin Hall ratio $\theta_{SH}$, spin diffusion length $\lambda_s$, spin mixing conductance $G_{\uparrow\downarrow}$, and damping due to spin memory loss $\alpha_{SML}$, for the Py/$Ni_{0.6}Cu_{0.4}$ bilayer. For this composition, the spin

Hall ratio reaches unity, $\theta_{SH} \approx 1$. Our self-consistent fitting method does not rely on parameters calculated from theory or taken from experiments on other materials. We compare our results with similar experiments on Py/Pt bilayers measured and analyzed using the same methods [14]. In the discussion section, we review several recent experiments on SHE in Pt that find much smaller values of $\theta_{SH}$ and we speculate on possible reasons for the differences among reported values.

## II.  ALLOY BAND STRUCTURE

The top plot in Figure 1 shows the band structure for fcc Ni without spin-polarization effects, i.e. for nonmagnetic Ni. These bands are quite similar to those of Pt (see Figure 11 in Appendix 1), especially where they cross the Fermi level near the L and X points. Although the calculated SOC parameter is an order of magnitude smaller for Ni than for Pt, it causes band splitting in nonmagnetic Ni analogous to that in Pt (see discussion and Figures 11 and 12 in Appendix 1). The bottom plot in Figure 1 shows the Bloch spectral function for $Ni_{60}Cu_{40}$, the alloy that is the main focus of our experiments. Comparing with the top plot shows that (1) the Fermi level is nearly unchanged, in contrast to what would be expected from a rigid band approximation, and (2) despite broadening due to disorder, the main features of nonmagnetic Ni can be recognized in the spectrum of $Ni_{60}Cu_{40}$, including those around the Fermi level near the L and X points. Thus if the splitting of degenerate bands near these points leads to an intrinsic spin Hall effect in nonmagnetic Ni as it does in Pt, then our calculations suggest it may also do so in $Ni_xCu_{1-x}$ alloys. Our theoretical methods are described in Appendix 1, along with a more detailed comparison of the effect of SOC in Pt and nonmagnetic Ni.

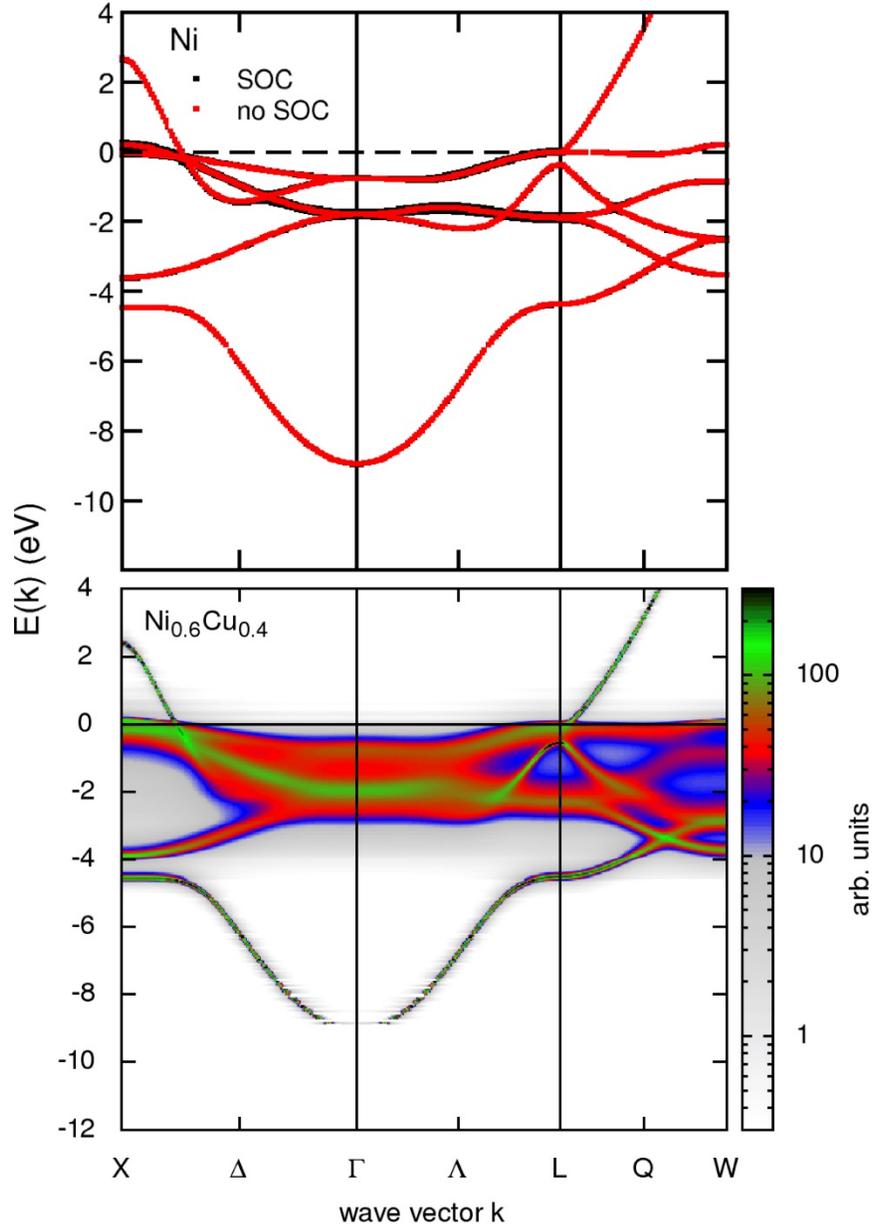

Figure 1 (color online): Band structure for nonmagnetic fcc Ni with and without SOC (top) and Bloch spectral function for $Ni_{60}Cu_{40}$ averaged over random alloy disorder configurations with SOC (bottom). The color scale for the bottom plot is proportional to occupation probability. Dashed horizontal line represents the Fermi level. See Appendix 1 for theoretical methods.

## III. EXPERIMENTAL TECHNIQUE

A FM/NM sample placed on a coplanar waveguide (CPW) presents an inductive load with real and imaginary components whose effect on CPW transmission can be measured in both amplitude and phase by a vector network analyzer (VNA). An overview of VNA-FMR measurements, including the underlying physics and practical considerations, can be found in [15]. Traditional VNA-FMR measurements record the change in amplitude at a fixed frequency $f$ as a function of an applied magnetic field $H_{app}$ that is swept through the resonance condition. The resonance field $H_{res}$ and linewidth $\Delta H$ measured over a range of $f$ yield several parameters (see, e.g., Section III of [16] for details): the Gilbert damping parameter $\alpha$, the Landé spectroscopic splitting factor $g$, the effective magnetization $M_{eff}$, and a measure of inhomogenous broadening $\Delta H_0$. For this traditional analysis, only the relative change in signal amplitude is used.

A recent advance in VNA-FMR exploits both the absolute amplitude and the phase of the VNA signal in order to extract additional information from the resonance [13]. This method relies on the fact that, at resonance, individual contributions to the total inductive signal from several phenomena in the FM/NM sample can be separated according to their phases relative to the driving ac magnetic field from the CPW, as well as by their dependence on $f$. As described in [13], the individual contributions consist of a real inductance $L_0$ and three complex inductances that can be conveniently described in terms of effective conductivities: (1) a Faraday term, $\sigma^F$, due to ac current in the NM generated via Faraday induction from the dipolar field of the precessing magnetization; (2) a fieldlike SOT term, $\sigma_{FL}^{SOT}$, due to forward and inverse REE or similar processes at the FM/NM interface, that appears together with $\sigma^F$ but can be separated as described below; and (3) a dampinglike SOT term, $\sigma_{DL}^{SOT}$, due to the linear superposition of forward and inverse SHE in the NM. A practical advantage of this method for detecting SOT effects is that is does not require lithographic patterning of the sample or direct electrical contacts to the sample.

The conductivities just described, like the ordinary charge conductivity $\sigma$, describe dissipative currents that flow in response to driving forces. In a generalized matrix form of Ohm's law [13], the diagonal elements of the conductivity matrix describe the usual response to an applied electric field and the Gilbert damping of the FM, while the off-diagonal elements describe how an electric field exerts torque on the FM and how the precessing magnetization of the FM generates charge current in the NM, through a combination of Faraday induction and SOT processes. Onsager reciprocity requires that the same conductivity describes both forward and inverse processes [4] of each type, which add together in phase for this measurement method [13].

1. Samples

Films were deposited via sputtering onto oxidized Si wafers (resistivity 10 Ω cm; oxide thickness 165 nm) at room temperature on a rotating sample holder in a vacuum chamber with a base pressure of $2.7 \times 10^{-7}$ Pa ( $2 \times 10^{-9}$ Torr ). The stack for samples without Py was substrate/Ta(3)/Ni$_x$Cu$_{1-x}$(20)/Ta(3),

where the parentheses show thickness in nm. The stack for samples with Py was substrate/Ta(3)/Py(3.5)/Ni$_x$Cu$_{1-x}$($d_{NM}$)/Ta(3). We deposited two series of samples with Py: (1) an alloy composition series with $d_{NM}$ = 10 nm and $x$ ranging from 0 to 1; (2) an alloy thickness series with $x$ = 0.6 and $d_{NM}$ ranging from 2 nm to 40 nm. The Ni$_x$Cu$_{1-x}$ layers were deposited by cosputtering from Ni and Cu targets with relative rates adjusted through the power supplies. Deposition rates were calibrated using X-ray reflectivity measurements.

After deposition, the wafers were coated with photoresist (≈ 1.5 µm thick) and then chips approximately 6 mm square were made via a scribe-and-break method. The dimensions of each chip were measured with calipers. The photoresist was stripped from chips used for magnetometry but left on chips used for FMR to avoid shorting the CPW.

### 1. Magnetometry

We measured magnetic moment $m$ vs. temperature $T$ for Ni$_x$Cu$_{1-x}$(20) samples using vibrating-sample magnetometry (VSM) or superconducting quantum interference device (SQUID) magnetometry. We fit $m$ vs. $T$ as described below to determine the FM transition temperature $T_c$. These measurements were done in a saturating magnetic field applied in the sample plane and the value of $T_c$ was independent of the applied field. We determined the saturation moment per unit area $m_A$ for Py/ Ni$_x$Cu$_{1-x}$(10) samples from in-plane hysteresis curves measured using SQUID magnetometry.

### 2. FMR Measurements

Samples were placed with the film side down onto a CPW mounted between the poles of an electromagnet capable of dc fields up to $\mu_0 H_0 \approx 2.2$ T, applied perpendicular to the plane of the sample. The center conductor of the CPW is 100 µm wide and the sample length along the center conductor was measured to allow an accurate comparison of absolute inductance among all samples [13]. After applying a saturating field, we used a VNA to record the real and imaginary parts of the transmission parameter $S_{21}$ at fixed $f$ as $H_0$ was swept through the FMR resonance. Our procedure for fitting the VNA data is described in detail in [14]. Key points we wish to highlight are: (1) The signal from the sample itself is de-embedded from effects due to cables, connectors, and the CPW that change with frequency. (2) Extraction of the usual spectroscopic parameters $H_{res}$, $\Delta H$, $\alpha$, $g$, $M_{eff}$, and $\Delta H_0$ is done as in traditional VNA-FMR (see, e.g., Section III of [16]). (3) The Faraday term $\sigma^F$ is separated from $\sigma_{FL}^{SOT}$, and both $\sigma_{FL}^{SOT}$ and $\sigma_{DL}^{SOT}$ are corrected for the fact that some of the induced current is shunted away from the CPW by the metallic sample. These last steps are particularly important for this study because the bulk resistivity of $Ni_xCu_{1-x}$ varies strongly with composition [17]. Details of this shunting correction are given in Appendix 2.

Each chip was measured three times to provide an estimate of reproducibility. Error bars shown in the plots below reflect one standard deviation uncertaintites from repeatability (variation of the three measurements of each sample) and quality of fit (uncertainty from the covariance matrix generated by the nonlinear least-squares fit).

## IV. SPIN TRANSPORT MODEL

Figure 2 shows various paths for spin current flow in a Ta/Py/NM sample. The total Gilbert damping of the Py can be expressed as a sum of three terms,

$$\alpha = \alpha_0 + \alpha_{sp}^{Py/NM}(d_{NM}) + \alpha_{SML}^{Py/NM}. \quad (1)$$

The first term includes all damping that does not involve the NM layer: Py intrinsic damping $\alpha_{int}^{Py}$ (which we expect to be close to the value of $\alpha_{int}^{Py} = 0.0054 \pm 0.0001$ we reported for Py/Cu control samples in [14]), possible spin memory loss (SML) at the Ta/Py interface, and spin pumping into the Ta seed layer. The second term is due to spin pumping into the NM layer and the third term represents SML at the Py/NM interface.

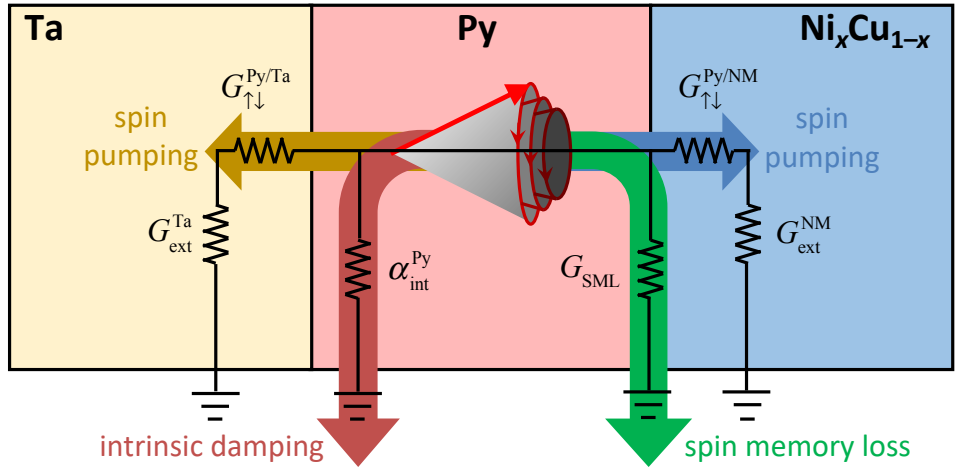

Figure 2 (color online): Schematic circuit diagram for angular momentum flow from the precessing magnetization of the Py layer to the various damping sources discussed in the text.

Damping due to spin pumping is proportional to an effective spin mixing conductance $G_{eff}^{sp}$ (units of $\Omega^{-1} m^{-2}$):

$$\alpha_{sp}(d_{NM}) = \left(\frac{\gamma \hbar^2}{2m_A e^2}\right) G_{eff}^{sp}, \tag{2}$$

where $\gamma = g\mu_B / \hbar$ is the gyromagnetic ratio of Py, $\hbar$ is the reduced Planck constant, $e$ is the elementary charge, $\mu_B$ is the Bohr magneton, and $m_A$ is the saturation moment per unit area. Following [18], $G_{eff}^{sp}$ is a series combination of the interfacial spin mixing conductance $G_{\uparrow\downarrow}$ and the thickness-dependent spin conductance $G_{ext}(d_{NM})$ within the NM:

$$\frac{1}{G_{eff}^{sp}} = \frac{1}{G_{\uparrow\downarrow}} + \frac{1}{G_{ext}} \quad \text{or} \quad G_{eff}^{sp} = \frac{G_{ext}}{1 + G_{ext}/G_{\uparrow\downarrow}}. \tag{3}$$

For a NM layer with bulk electrical conductivity $\sigma_0$ and spin diffusion length $\lambda_s$,

$$G_{ext}(d_{NM}) = (\sigma_0/2\lambda_s)\tanh(d_{NM}/\lambda_s). \tag{4}$$

For the Ni$_{60}$Cu$_{40}$ alloy used in this work, $\sigma_0$ was measured to be independent of NM thickness (see Section V.7) and we assume $\lambda_s$ is also independent of NM thickness.

As illustrated in Figure 2, we treat SML as a conductance $G_{SML}$ in parallel with the spin pumping conductance, giving a total effective conductance of

$$G_{eff}^{tot} = G_{eff}^{sp} + G_{SML} = \frac{G_{ext}}{1 + G_{ext}/G_{\uparrow\downarrow}} + G_{SML}. \tag{5}$$

Spin current flowing through $G_{SML}$ does not enter the NM and therefore cannot contribute to the SHE signal. We define an efficiency $\varepsilon$ as the fraction of the total spin current that does enter the NM,

$$\varepsilon = \frac{G_{eff}^{sp}}{G_{eff}^{sp} + G_{SML}} = \frac{\alpha_{sp}}{\alpha_{sp} + \alpha_{SML}}, \tag{6}$$

normally taken in the limit $d_{NM} \gg \lambda_s$ to obtain an efficiency that is independent of NM layer thickness. The fraction of spin current lost at the interface is simply $1-\varepsilon$.

Combining Equations (2) through (5), we can rewrite Equation (1) to make the various parameters explicit

$$\alpha(d_{NM}) = \alpha_0 + \left(\frac{\gamma \hbar^2}{2m_A e^2}\right)\left[\frac{(\sigma_0/2\lambda_s)\tanh(d_{NM}/\lambda_s)}{1+[(\sigma_0/2\lambda_s)/G_{\uparrow\downarrow}]\tanh(d_{NM}/\lambda_s)} + G_{SML}\right]. \quad (7)$$

The spin current pumped into the NM layer is converted to a charge current by the inverse spin Hall effect (iSHE), which is precisely the process described by the dampinglike SOT conductivity $\sigma_{DL}^{SOT}$. We use a result derived in [19] to express $\sigma_{DL}^{SOT}$ in terms of the spin Hall conductivity $\sigma_{SH}$ of the NM layer (related to the spin Hall ratio by $\theta_{SH} \equiv \sigma_{SH}/\sigma_0$) and the various interfacial and NM parameters:

$$\sigma_{DL}^{SOT}(d_{NM}) = \sigma_{SH}\varepsilon\left[\frac{(1-e^{-d_{NM}/\lambda_s})^2}{(1+e^{-2d_{NM}/\lambda_s})} \frac{|\tilde{G}_{\uparrow\downarrow}|^2 + \text{Re}\{\tilde{G}_{\uparrow\downarrow}\}\tanh^2(d_{NM}/\lambda_s)}{|\tilde{G}_{\uparrow\downarrow}|^2 + 2\text{Re}\{\tilde{G}_{\uparrow\downarrow}\}\tanh^2(d_{NM}/\lambda_s) + \tanh^4(d_{NM}/\lambda_s)}\right] \quad (8)$$

where $\tilde{G}_{\uparrow\downarrow} \equiv G_{\uparrow\downarrow}(2\lambda_s/\sigma_0)\tanh(d_{NM}/\lambda_s)$. Thus the measured $\sigma_{DL}^{SOT}$ is proportional to $\sigma_{SH}$ of the NM layer, the efficiency $\varepsilon$ of the FM/NM interface, and a term in square brackets that depends on NM thickness as well as spin transport parameters at the FM/NM interface and in the NM bulk. This last term accounts for the boundary condition that the spin current vanishes at the far surface of the NM layer [19].

## V. EXPERIMENTAL RESULTS

### 1. Curie Temperature for Composition Series

The decrease in saturation magnetization $M_s$ as $T$ approaches $T_c$ from below is predicted by continuous phase transition theory to be described by a power law,

$$M_s \propto (T_c - T)^{-\beta}, \quad (9)$$

where the critical exponent $\beta$ is predicted to be 0.5 from mean field theory [20]. Figure 3(a) shows a measured $m$ vs $T$ curve for $Ni_{70}Cu_{30}$ and a fit using Eqn. (9) that gives $T_c = (266 \pm 2)$ K and $\beta = 0.37 \pm 0.02$. (All uncertainties here represent one standard deviation.) Exponents smaller than 0.5 have been reported for Ni, Fe, and other FM metals [20].

Figure 3(b) shows $T_c$ vs Ni fraction for $x \geq 0.55$. For smaller values of $x$, $T_c$ was too small to determine reliably from $m$ vs $T$. The dashed curve shows values for bulk $Ni_xCu_{1-x}$ alloys, taken from Fig. 8-51 of [12]. These data show that compositions with $x \leq 0.7$ are candidates for a detailed study of SHE in the paramagnetic version of the alloy at 300 K. Our focus was narrowed further by a survey of FMR properties, which we describe next.

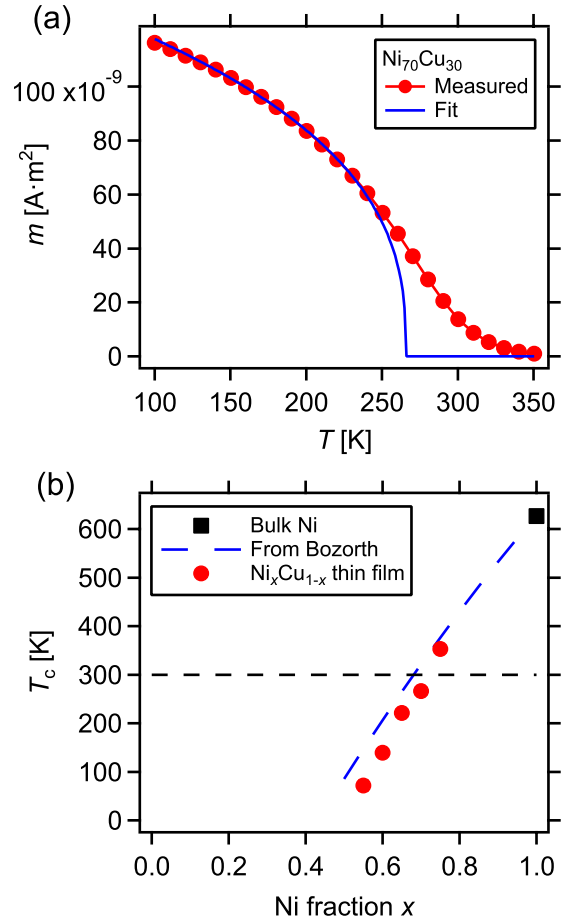

Figure 3: (a) Saturation magnetic moment vs temperature for $Ni_{70}Cu_{30}(20)$. Solid curve is a fit using Eqn. (9), giving $T_c$ = 266 K and $\beta$ = 0.37. Data above $T$ = 230 K were excluded from the fit to avoid artifacts due to the divergence of the paramagnetic susceptibility near $T_c$. (b) Curie temperature for $Ni_xCu_{1-x}$ alloys. Horizontal dashed line indicates room temperature. Dashed curve for bulk samples is from Fig. 8-51 of [12].

2. Moment Per Unit Area for Composition Series

Figure 4 shows the moment per unit area, $m_A$, for $Py(3.5)/Ni_xCu_{1-x}(10)$ measured with an in-plane applied field of 1.6 kA/m (20 Oe). For $x \geq 0.5$, $m_A$ rises slowly due to the induced moment in the $Ni_xCu_{1-x}$ layer. The value for $x = 0$, i.e. for $Py/Cu(10)$, can be used to estimate the active thickness $d_{Py}$ of the Py layer using $m_A = M_s d_{Py}$. From $m_A = (2.18 \pm 0.1)$ mA at $x = 0$, and assuming the nominal value $M_s$ = 800 kA/m for our Py films, we find $d_{Py} = (2.72 \pm 0.14)$ nm, implying a dead layer thickness of $(0.78 \pm 0.14)$ nm. This is consistent with previous reports of dead layer thickness for Py deposited on a Ta seed layer [21].

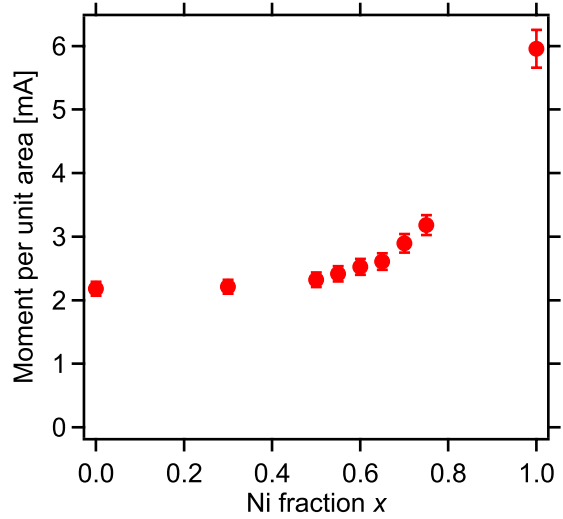

Figure 4: Moment per unit area for Py(3.5)/Ni$_x$Cu$_{1-x}$(10) samples with an in-plane applied field of 1.6 kA/m (20 Oe).

## 3. Spectroscopic FMR Parameters for Composition Series

Figure 5 shows FMR data for Py(3.5)/Ni$_x$Cu$_{1-x}$(10) samples spanning the full composition range. For $x \leq 0.6$, where the alloy is paramagnetic, the increase in $\alpha$ with Ni fraction is expected because the spin current generated by spin pumping is absorbed much more efficiently by Ni than by Cu. For larger $x$, exchange coupling between the Py layer and the weakly ferromagnetic alloy layer leads to a further increase in damping. The increase in $g$ and the decrease in $M_{eff}$ for $x > 0.5$ are both consistent with an increase in perpendicular magnetic anisotropy of the Py with increasing Ni fraction in the alloy [22]. The small and constant value of $\Delta H_0$ indicates the magnetic properties of the films are spatially uniform across the full composition range.

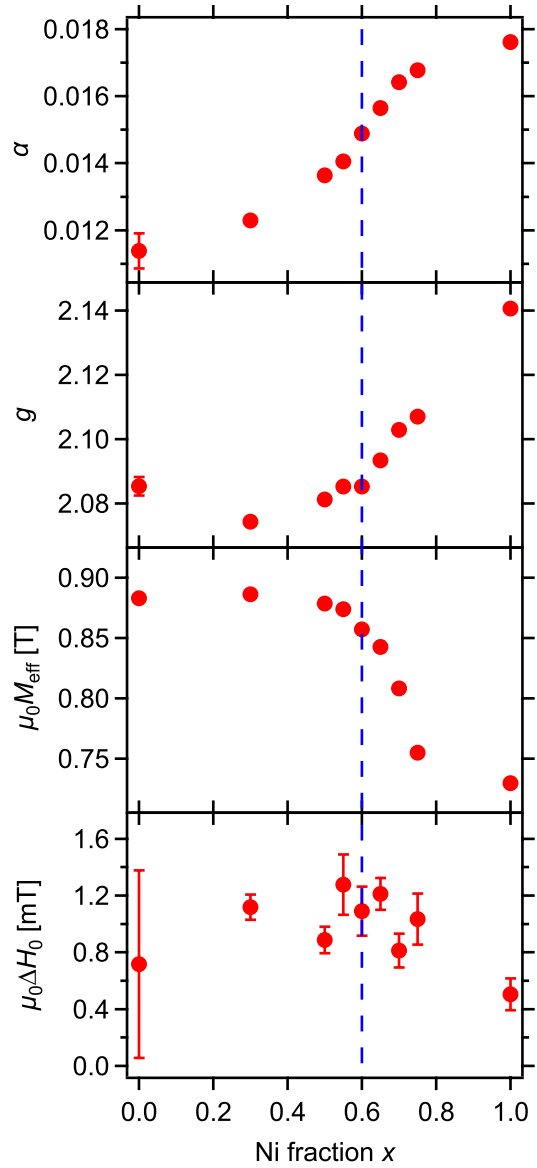

Figure 5: Spectroscopic FMR parameters for Py(3.5)/Ni$_x$Cu$_{1-x}$(10) samples. Vertical dashed line indicates the $x = 0.6$ composition used to determine the microscopic spin transport parameters given in Table 1.

## 4. SOT Conductivities for Composition Series

The top plot in Figure 6 shows $\sigma_{DL}^{SOT}$ for the Py(3.5)/Ni$_x$Cu$_{1-x}$(10) samples. As described above, this conductivity directly measures the process whereby spin current pumped from the FM is converted by the iSHE to charge current in the NM layer (which affects $S_{21}$ via the Oersted field at the CPW), as well as the reciprocal process. The signal is largest for alloys near $x = 0.7$, where $T_c$ is near room temperature and the alloy is mostly paramagnetic. A similar enhancement of DL SOT near $T_c$ has been reported for Fe$_x$Pt$_{1-x}$ and attributed to spin fluctuations [23]. The $x = 0.6$ composition has a large SHE and is fully paramagnetic at room temperature, so we focus on this for the thickness series discussed in the next section.

The values of $\sigma_{DL}^{SOT}$ for pure Ni and pure Cu are shown for completeness, but there are caveats to note with these points. First, the pure Ni sample is a FM/FM bilayer for these room temperature measurements. There is recent evidence that the distinction between FL and DL torques for the FM/FM case may not be as clear as for the FM/NM case [24]. Second, the quality of fit for the pure Cu sample with $d_{NM} = 10$ nm was poor compared to the rest of the composition series, not only for $\sigma_{DL}^{SOT}$ but also for $\alpha$ and $\Delta H_0$ (see Figure 5). This was not the case for pure Cu samples with $d_{NM} \leq 2$ nm, so we suspect that as the Cu film grows thicker and rougher [25] it induces some change in the Py itself or in the Py/Cu interface. Finally, because the resistivity is much lower for the pure elements than for $x$ near 0.5 [17], the shunting correction for these two points is much larger than for the other points and removal of the Faraday contribution is therefore less accurate (see Appendix 2). For these reasons, quantitative conclusions cannot be drawn from the pure Ni and Cu samples without further investigation that is beyond the scope of this paper.

The bottom plot in Figure 6 shows $\sigma_{FL}^{SOT}$ for the Py(3.5)/Ni$_x$Cu$_{1-x}$(10) samples, excluding the pure Ni and pure Cu samples for the reasons just given. As with $\sigma_{DL}^{SOT}$, the signal is largest for alloys with $T_c$ near room temperature. The fact that the two SOT conductivities have comparable amplitudes is a coincidence of the choice of 10 nm as the alloy thickness for this composition series. As shown below for the thickness series with $x = 0.6$, $\sigma_{FL}^{SOT}$ has saturated at $d_{NM} = 10$ nm while $\sigma_{DL}^{SOT}$ is still increasing.

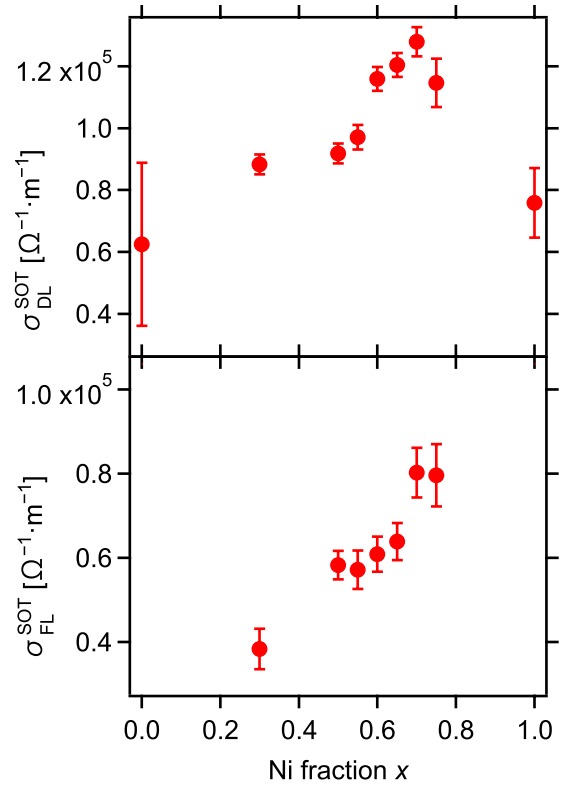

Figure 6: Dampinglike SOT conductivity (top) and fieldlike SOT conductivity (bottom) for Py(3.5)/Ni$_x$Cu$_{1-x}$(10) samples.

5. Spectroscopic FMR Parameters for Thickness Series with *x* = 0.6

Figure 7 shows FMR data for Py(3.5)/Ni$_{60}$Cu$_{40}$($d_{Ni60}$) samples with $d_{Ni60}$ ranging from 2 nm to 40 nm. All quantities are independent of thickness for $d_{Ni60} \geq 5$ nm. We attribute the changes for the thinner films to proximity effects that induce weak magnetic ordering in the first few nm of the otherwise nonmagnetic Ni$_{60}$Cu$_{40}$ (hereafter "Ni60") film. In particular, a similar linear dependence of $\alpha$ on NM thickness was observed for Py/Pt and Py/Pd bilayers in which proximity-induced magnetic moments were confirmed using X-ray measurements [26].

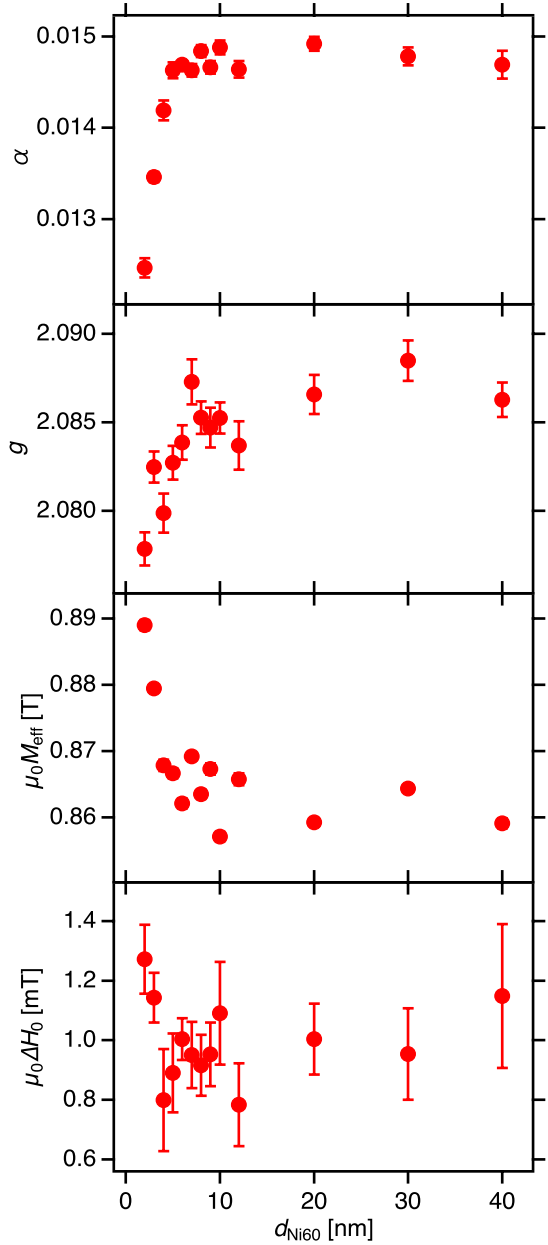

Figure 7: Spectroscopic FMR parameters for Py(3.5)/Ni$_{60}$Cu$_{40}$($d_{Ni60}$) samples.

## 6. SOT Conductivities for Thickness Series with *x* = 0.6

Figure 8 shows $\sigma_{DL}^{SOT}$ and $\sigma_{FL}^{SOT}$ for the Py(3.5)/Ni$_{60}$Cu$_{40}$($d_{Ni60}$) thickness series. The small contribution to $\sigma_{DL}^{SOT}$ from the Ta seed layer has been subtracted by measuring a control sample, Ta(3)/Py(3.5)/Cu(1)/Ta(3), whose value was $\sigma_{DL}^{SOT} = (2987 \pm 1422)$ $\Omega^{-1}$ m$^{-1}$. (Removal of the Ta seed contribution from $\sigma_{FL}^{SOT}$ is not as simple because of the Faraday contribution discussed in Appendix 2, but we do not attempt a quantitative analysis of the fieldlike signal in this work.) The monotonic increase in $\sigma_{DL}^{SOT}$ with NM thickness is similar to the dependence of SHE signal on Pt thickness in FM/Pt bilayers measured by various techniques [14, 27, 28]. In contrast, $\sigma_{FL}^{SOT}$ increases roughly linearly with NM thickness before saturating for $d_{Ni60} > 9$ nm. This is similar to the behavior seen for Py/Pt bilayers [14], although in that case $\sigma_{FL}^{SOT}$ was already saturated for $d_{Pt} = 2$ nm.

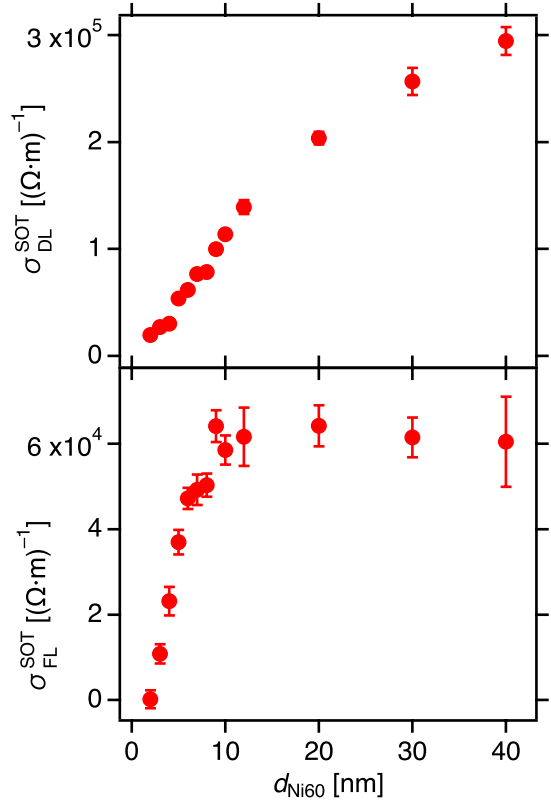

Figure 8: Dampinglike SOT conductivity (top) and fieldlike SOT conductivity (bottom) for Py(3.5)/Ni$_{60}$Cu$_{40}$($d_{Ni60}$) samples. The contribution from the Ta seed layer has been subtracted from $\sigma_{DL}^{SOT}$ (see main text) and the Faraday contribution has been removed from the inductive signal that contains $\sigma_{FL}^{SOT}$ (see Appendix 2). Both conductivities have been corrected for shunting (see Appendix 2).

## 7. Extracting Spin Transport Parameters

Following the approach used for Py/Pt reported in [13] and [14], we can extract microscopic spin transport parameters from the thickness series for $x = 0.6$. We discuss only the SHE-related parameters that can be extracted from $\sigma_{DL}^{SOT}$ because fewer assumptions are required than for extracting REE-related parameters from $\sigma_{FL}^{SOT}$ (see Section V of [13]). As discussed in Section III.C of [14], we enforce Onsager reciprocity on the processes of spin pumping and SHE by requiring that both are described by the same spin diffusion length. The input data for this approach comprise $\sigma_{DL}^{SOT}$ and $\alpha$ as function of NM thickness $d_{Ni60}$, which are replotted in Figure 9.

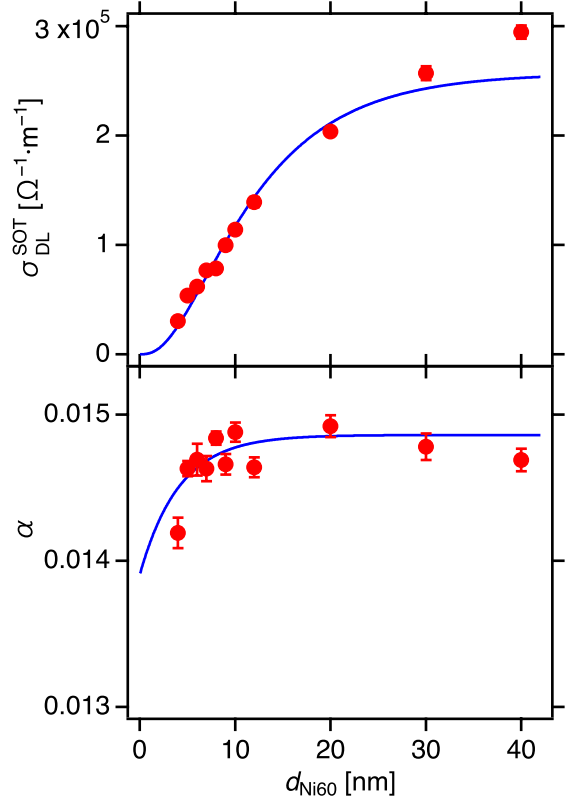

Figure 9: Fits to dampinglike SOT conductivity (top) and damping (bottom) using Equations (7) and (8). The points for $d_{Ni60}$ = 2 nm and 3 nm have been omitted because they are likely affected by proximity induced ferromagnetism (see main text).

The solid curves in Figure 9 are fits using Equations (7) and (8). The fixed parameter $\alpha_0$ was measured using a control sample, Ta(3)/Py(3.5)/Cu(1)/Ta(3), whose damping was $\alpha_0 = 0.01114 \pm 0.0001$. As described in Section III, $\alpha_0$ represents all damping that does not involve the NM layer. In using this control sample, we ignore SML at the Py/Cu interface, expected to be small because Cu has low SOC, and we assume the partially oxidized Ta cap layer does not contribute a significant amount of damping. This second assumption is by no means obvious, but the results in [14] for control samples with an oxide layer separating the Py from the Ta cap layer indicate the cap contributes no more than 0.001 to the total damping. We note that if we have underestimated damping by using this control sample it would imply less spin current entering the NM layer and therefore require *larger* values of $\sigma_{SH}$ and $\theta_{SH}$ to fit the data in Figure 9.

The remaining fixed parameters in Equations (7) and (8), $g$, $m_A$, and $\sigma_0$, were determined as follows. The mean value of $g$ from Figure 7, excluding the points for $d_{Ni60} < 5$ nm affected by proximity induced magnetism, is $g = 2.0854 \pm 0.0006$. The saturation moment per unit area $m_A$ was nearly constant for all films in the Ni60 thickness series, with a mean value of $m_A = (2.42 \pm 0.02)$ mA. Thus the factor in Equation (2) used to convert between damping and conductance per unit area is $\gamma \hbar^2 / 2 m_A e^2 = (16.4 \pm 0.14) \times 10^{18}$ $\Omega$ m$^2$. The bulk electrical conductivity $\sigma_0$ was determined from the sheet resistance $R_s$ measured for each Ni60 sample. We find that $1/R_s$ varies linearly with $d_{Ni60}$, as shown in Figure 10, so we fit the data using $1/R_s = \sigma_0^{Ni60} d_{Ni60} + 1/R_{other}$. This yields $\sigma_0^{Ni60} = (1.93 \pm 0.02) \times 10^6$ $\Omega^{-1}$m$^{-1}$ and a parallel sheet resistance from the Py and Ta layers of $R_{other} = (91 \pm 3)$ $\Omega$. The fact that $\sigma_0^{Ni60}$ is independent of thickness means our results here, unlike those reported for Pt in [14], [29], and [28], do not depend on a particular model for the spin-relaxation process or on whether the origin of the SHE in Ni$_{60}$Cu$_{40}$ is intrinsic or extrinsic.

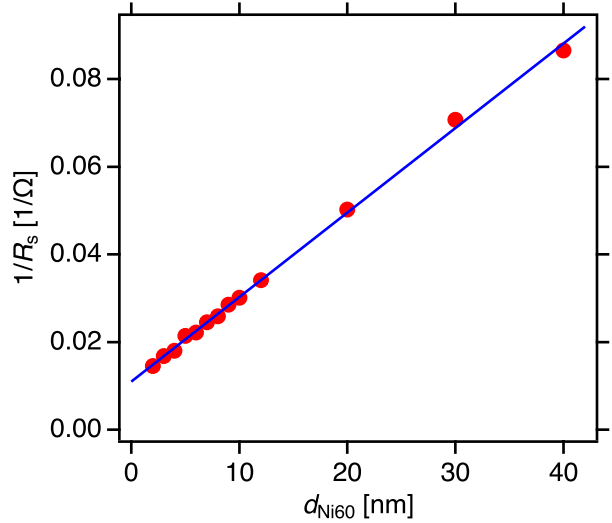

Figure 10: Inverse sheet resistance for Py(3.5)/Ni$_{60}$Cu$_{40}$($d_{Ni60}$) samples.

There are four remaining parameters in Equations (7) and (8) that are adjusted to give the fits in Figure 9: $G_{\uparrow\downarrow}$, $G_{SML}$, $\lambda_s$, and $\sigma_{SH}$. The results are given in Table 1, along with the results for Py/Pt bilayers reported in [14]. The table also shows the related quantities $\theta_{SH}$, $\varepsilon$, and $\alpha_{SML}$.

Table 1: Parameters extracted from fits shown in Figure 9. Values for Py/Pt are from [14].

|  | $G_{\uparrow\downarrow}$ | $G_{SML}$ | $\lambda_s$ | $\sigma_{SH}$ | $\theta_{SH} = \sigma_{SH}/\sigma_0$ | $\varepsilon$ | $\alpha_{SML}$ |
|---|---|---|---|---|---|---|---|
|  | [$10^{15}$ $\Omega^{-1}$ m$^{-2}$] | [$10^{15}$ $\Omega^{-1}$ m$^{-2}$] | [nm] | [$10^6$ $\Omega^{-1}$ m$^{-1}$] |  |  |  |
| Py/Ni$_{60}$Cu$_{40}$ | 0.115 ± 0.05 | 0.17 ± 0.10 | 8.3 ± 0.4 | 2.02 ± 0.35 | 1.05 ± 0.18 | 0.26 ± 0.09 | 0.0028 ± 0.0016 |
| Py/Pt | 1.3 ± 0.2 | 0.79 ± 0.22 | 4.2 ± 0.1 | 2.36 ± 0.04 | 0.387 ± 0.008 | 0.37 ± 0.06 | 0.012 ± 0.004 |

## VI. DISCUSSION

In terms of the conductivity $\sigma_{SH}$, which reflects the current induced by a unit excitation for both spin-to-charge (iSHE) and charge-to-spin (SHE) conversion, the SHE in Py/Ni$_{0.6}$Cu$_{0.4}$ is nearly equal to that in Py/Pt. The much larger value of $\theta_{SH}$ for Ni60 follows from the fact that $\sigma_0$ for Ni60 is much smaller than for Pt ($\sigma_0^{Pt} = (6.13 \pm 0.04) \times 10^6$ 1/($\Omega$ m) [14]). This illustrates the importance of $\sigma_0$ for the efficiency of spin-charge interconversion, and the potential advantage for applications of alloy SHE materials in which $\sigma_0$ may be tunable somewhat independent of $\sigma_{SH}$. Although a spin Hall ratio greater than unity may seem surprising, it is possible due to the transverse nature of the interconversion process in the SHE, as explained in Section II of [3].

Comparing $G_{SML}$ for Ni60 and Pt, we see it is about 5x smaller in Ni60. This is not surprising since SML is associated with SOC, which is much smaller in Cu and Ni than in Pt [30]. Less expected is the fact that $G_{\uparrow\downarrow}$ is about 10x smaller in Ni60 than in Pt, indicating a much less transparent interface when the former is deposited onto Py. Calculated values [31] for Py/Pt, $G_{\uparrow\downarrow} = 1.07 \times 10^{15}$ $\Omega^{-1}$ m$^{-2}$, and Py/Cu, $G_{\uparrow\downarrow} = 0.48 \times 10^{15}$ $\Omega^{-1}$ m$^{-2}$, are both within about 10 % of the theoretical upper limit given by the Sharvin conductance $G_{Sh}$ [31], and experimental $G_{\uparrow\downarrow}$ values for Py/Pt (Table 1) and Py/Cu [32] are within about 20 % of the calculated values. The value of $G_{\uparrow\downarrow}$ for Ni60 in Table 1 is more than 4x smaller than the calculated $G_{\uparrow\downarrow}$ of Py/Cu, but we are not aware of any calculated values of $G_{\uparrow\downarrow}$ or $G_{Sh}$ for pure Ni or Ni-Cu alloys (or any alloys at all) to which this value can be compared.

The spin diffusion length $\lambda_s$ is about twice as long in Ni60 as in Pt, which is again consistent with smaller SOC in the alloy. We note that as $\theta_{SH}$ becomes large, the diffusion of spins in the NM will be limited by spin-charge conversion as well as by the usual spin-flip scattering. Combined with the fact that the mean free path in many materials, including Pt, is not much smaller than $\lambda_s$, this is further reason to consider the values of $\theta_{SH}$ and $\lambda_s$ as phenomenological parameters rather than bulk values for the NM material (see e.g., Section III.D of [4]).

From the efficiency $\varepsilon$, we see that SML plays a somewhat larger role in Py/Ni60 than in Py/Pt: the fraction of spin current lost to SML, $1-\varepsilon$, is 74 % for Py/Ni60 and 63 % for Py/Pt. There is potential for significant improvement in both cases if the source of SML can be identified and reduced.

We now want to put the results in Table 1, particularly the large values of $\theta_{SH}$, in the broader context of other SHE experiments. Several papers have noted the wide variation in reported values for $\theta_{SH}$ and related parameters of a given NM material [4, 27, 29]. Given the different experimental techniques and the different models used to extract these parameters, multiple effects contribute to these widely varying results. One of the most important effects is that interfacial SML was not taken into account in many early studies. This limitation results in an underestimate of $\theta_{SH}$, regardless of the experimental technique employed, and the size of the underestimate can be a factor of two or even larger. In addition, values for $G_{\uparrow\downarrow}$ and $\lambda_s$ that are assumed rather than determined directly in the experiments can strongly affect the value of $\theta_{SH}$ inferred from a particular experiment. Measurements spanning a range of NM thickness are particularly helpful in this respect.

While a detailed comparison with other results for SHE is beyond the scope of this paper, we can comment on specific differences in approach that may contribute to our value of $\theta_{SH}$ being much larger than that reported by other groups. We focus on studies of Pt, but the comments apply to other materials as well. We first consider experiments that do not include SML and then move to those that do.

A series of studies by a group at Cornell [28, 33, 34] has progressively improved both the experimental techniques and the spin transport model for FM/Pt bilayers. The ratio of symmetric to antisymmetric components of the spin-torque FMR peak, the method used in [33] and [34], does not include the fieldlike SOT and thereby underestimates $\theta_{SH}$, so [34] varied the FM thickness in order to compensate for this omission. An analysis of damping vs. FM thickness to directly determine $G_{\uparrow\downarrow}$ yielded $\theta_{SH} = 0.33 \pm 0.05$ for unannealed CoFe/Pt(4) samples but gave unphysical results for three other sample types [34]. This value agrees with the value in Table 1, but it must be noted that it is based on assuming a very small value for the

Pt spin diffusion length, $\lambda_s = 1.4$ nm. Later work based on 2nd-harmonic anomalous Hall voltages [28] varied the Pt thickness in order to directly determine $\lambda_s$ for bulk Pt, obtaining a value of $\lambda_s = (5.1 \pm 0.5)$ nm, but this work did not vary the FM thickness and therefore assumed a calculated value for $G_{\uparrow\downarrow}$. Although [28] does not give an explicit value for $\theta_{SH}$, the reported values of $\sigma_{SH}$ and $\sigma_0$ can be combined to give $\theta_{SH} = \sigma_{SH}/\sigma_0 = 0.089 \pm 0.003$. The authors of [28] acknowledge that this value is a lower bound due the absence of SML in their model. In fact, the quantity $\xi_{DL}^E$ in Figure 2(c) of [28] can be directly compared with $\sigma_{DL}^{SOT}$ in Figure 8(a) of [14], and for the thickest Pt films the two quantities are quite close: $\xi_{DL}^E \approx 2.8 \times 10^5$ $\Omega^{-1}$m$^{-1}$ while $\sigma_{DL}^{SOT} \approx 4.9 \times 10^5$ $\Omega^{-1}$m$^{-1}$. Thus the phenomenological magnitude of the effect in [28] and [14] is similar, but the inclusion of SML in the model of [14] leads to a much larger value of $\theta_{SH}$.

Turning to experiments on Pt in which SML was included in the models, we consider [27] and [29]. Both of these studies combine two measurements, inverse SHE voltage induced in the Pt and change in damping between FM/Pt samples and an FM-only reference sample. A Pt thickness series allowed for a direct determination of $\lambda_s$ in both cases. The inclusion of SML differed in the two cases: [27] adopted results from several magnetoresistance studies at $T = 4$ K to estimate interfacial spin resistances and losses that are fixed parameters in a microscopic three-layer model of FM/NM spin transport, while [29] used a phenomenological parameter, equivalent to our $\varepsilon$, and took this as an adjustable parameter in fitting the experimental data. The resulting SML for Co/Pt interfaces was $\varepsilon = 0.41$ in [27] and $\varepsilon = 0.61$ in [29]. Reference [29] also reported $\varepsilon = 0.37$ for the Py/Pt interface. These two studies, along with the values of $\varepsilon$ in Table 1, support the view that a significant fraction of spin current is lost at metallic FM/NM interfaces.

Despite the inclusion of SML in the models, both [27] and [29] report $\theta_{SH}$ values that are quite small: $\theta_{SH} = 0.056 \pm 0.01$ [27] and $\theta_{SH} = 0.029 \pm 0.001$ [29] for Co/Pt, and $\theta_{SH} = 0.032 \pm 0.004$ [29] for Py/Pt. We can offer only speculative reasons for these surprisingly small values. In the case of [27], the iSHE voltage is proportional to $H_{rf}^2$, where $H_{rf}$ is the ac magnetic field at the FM layer of the sample. Although the microwave field in the cavity was calibrated in [27], numerical evaluation of Maxwell's equations for multilayer samples with appropriate boundary conditions [35] shows that highly conductive layers such as Pt can significantly reduce $H_{rf}$ at the FM layer itself, well beyond the amount predicted by a simple skin depth analysis. If $H_{rf}$ was one third the expected value due to such shielding effects, the same iSHE voltage would correspond to a value of $\theta_{SH}$ roughly 9 times larger. In the case of [29], the expression used to fit $\alpha(d_{NM})$ is unconventional in two significant ways, as discussed in Appendix 3. The result is that, compared to the model used here and in [14], the model used in [29] predicts a much larger fraction of the spin current generated by the FM enters the NM where it can contribute to the SHE signal, so it yields a much smaller value of $\theta_{SH}$.

## VII. CONCLUSION

In summary, the alloy $Ni_xCu_{1-x}$ shows a dampinglike SOT conductivity comparable to that of Pt for a wide range of composition. For $x = 0.6$, where the alloy is paramagnetic at room temperature, we found a spin Hall ratio near unity and an interfacial spin memory loss near 75 %. If interfacial loss can be understood and reduced, this alloy system offers advantages for applications of SOT, including the possibility that the spin Hall and bulk charge conductivities can be separately tuned by adjusting composition. Although our experimental results cannot distinguish between intrinsic and extrinsic mechanisms for the SHE in $Ni_xCu_{1-x}$, our ab initio band structure calculations show features resembling those underlying the intrinsic SHE in Pt, suggesting there may also be a significant intrinsic effect in the alloy.

## VIII. CONTRIBUTIONS

TJS and MWK designed the experiments. KSG, MWK, and TJS performed the measurements and data analysis. MA and JMS deposited the multilayer films and JMS performed the X-ray reflectivity measurements. EKDC performed the band structure and Bloch spectral function calculations. MWK wrote the manuscript and all authors reviewed and edited the manuscript.

## IX. ACKNOWLEDGMENTS

We are grateful to Mark Stiles for pointing out the similarity between the Fermi surfaces of Pt and nonmagnetic Ni, and to Andrew Berger and Mark Stiles for valuable comments on the manuscript. EKDC acknowledges the Swedish Research Council (VR) Grants No. 2016-04524, No. 2016-06955, and No. 2013-08316, STandUP and eSSENCE for financial support and the Swedish National Infrastructure for Computing (SNIC) for computational resources.

## APPENDIX 1: THEORETICAL METHODS

The band structures of Pt, Ni, and $Ni_xCu_{1-x}$ were calculated within the framework of density functional theory [36, 37]. The Kohn-Sham equations were solved as implemented in the spin-polarized relativistic Korringa-Kohn-Rostoker (SPR-KKR) code [38, 39]. For the alloys, random chemical disorder was treated within the coherent potential approximation (CPA) [40, 41]. We used the Vosko-Wilk-Nusair [42] version of the local spin density approximation for the exchange-correlation functional. The shape of the potential was considered by using the atomic sphere approximation for both spin-polarized and non-spin-polarized calculations. The effect of SOC was determined by comparing band structures obtained using the scalar relativistic approximation with those using the full relativistic Dirac equation.

We sampled the irreducible wedge of the Brillouin zone with 1500 $k$-points and included $s, p, d,$ and $f$ orbitals in the basis set ($l_{max} = 4$). The lattice parameters for fcc Pt, Ni and Cu taken from [43] are 3.923 Å, 3.523 Å and 3.614 Å, respectively. The lattice parameters for the alloys estimated using Vegard's law are 3.560 Å for $Ni_{0.6}Cu_{0.4}$ and 3.542 Å for $Ni_{0.8}Cu_{0.2}$.

The calculated total SOC parameter at the Fermi energy, $\xi(E_F)$, is 1.65 eV for Pt and about an order of magnitude smaller for both Ni (0.17 eV) and Cu (0.29 eV). The calculated $d$-orbital contributions to $\xi(E_F)$, 0.71 eV for Pt, 0.11 eV for Ni, and 0.14 eV for Cu, agree with values reported previously [44]. The average $\xi(E_F)$ for $Ni_{0.6}Cu_{0.4}$ alloy, 0.28 eV, is larger than that of Ni because $\xi(E_F)$ is larger for Cu than for Ni.

Figure 11 and Figure 12 show the effect of SOC on the bands of Pt and fcc Ni without spin-polarization effects. The Ni bands are less broad overall because $3d$ wave functions overlap less than $5d$ wave functions, but otherwise the Ni and Pt bands are quite similar. The splitting of degenerate bands by SOC is evident throughout the Brillouin zone of Pt, including near the L and X points where bands cross the Fermi level. These results are in agreement with those presented in [10]. The SOC splittings are much smaller in Ni, but the expanded view of bands near the Fermi level shown in Figure 12 reveals they are qualitatively similar to those in Pt. In particular, the degeneracy of both bands near X is lifted and the bands that cross the Fermi level near L not only split but also change curvature.

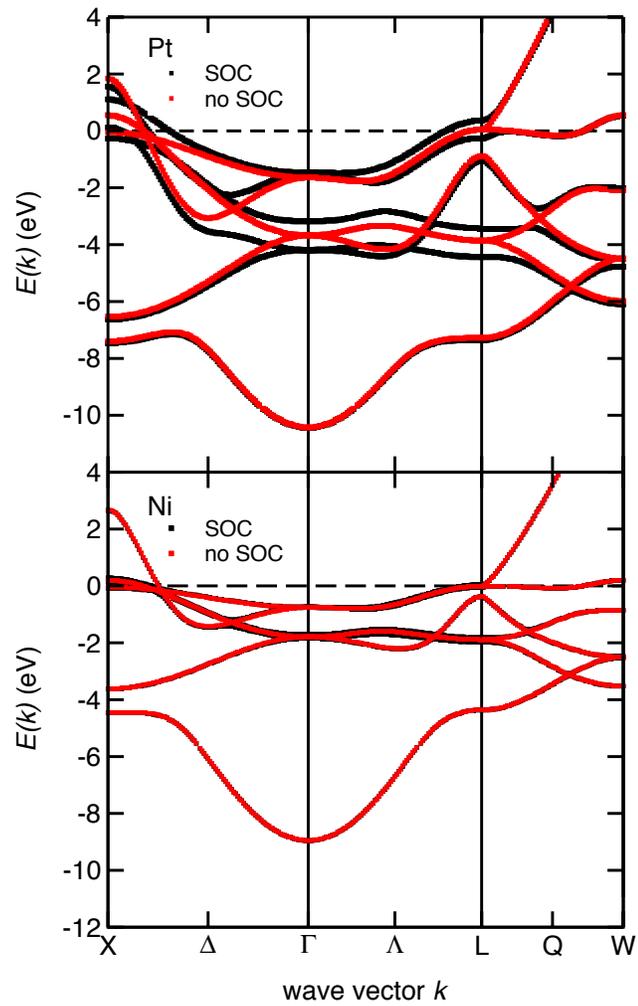

Figure 11: Band structure of Pt (top) and fcc Ni (bottom) without spin-polarization effects. Black points are with SOC and red points are without SOC. Dashed horizontal line represents the Fermi level.

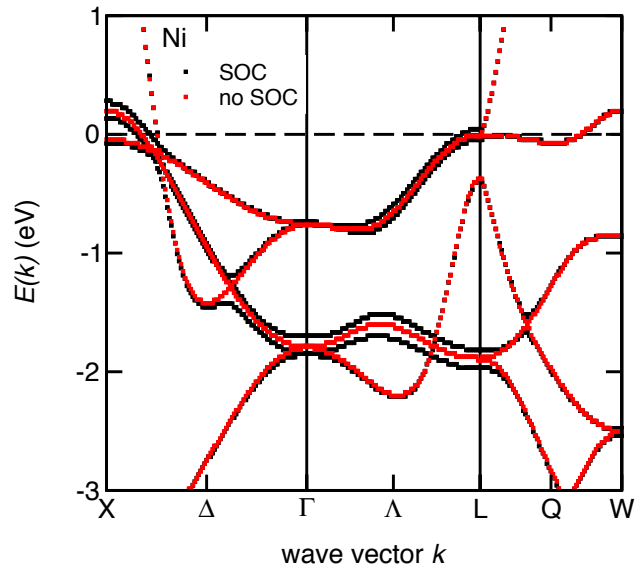

Figure 12: Expanded view of band structure for fcc Ni without spin-polarization effects, showing the splitting due to SOC of degenerate bands at the Fermi level near the L and X points. Black points are with SOC and red points are without SOC.

As shown in Figure 13, the band structure of ferromagnetic Ni does not resemble that of Pt, especially near the important L and X points. The Ni bands near X are split due to spin-polarization, but there is no further splitting due to SOC. There is also a small splitting due to SOC along Γ-L, but the curvature of this band where it crosses the Fermi level, between Λ and L, does not change as it does for nonmagnetic Ni.

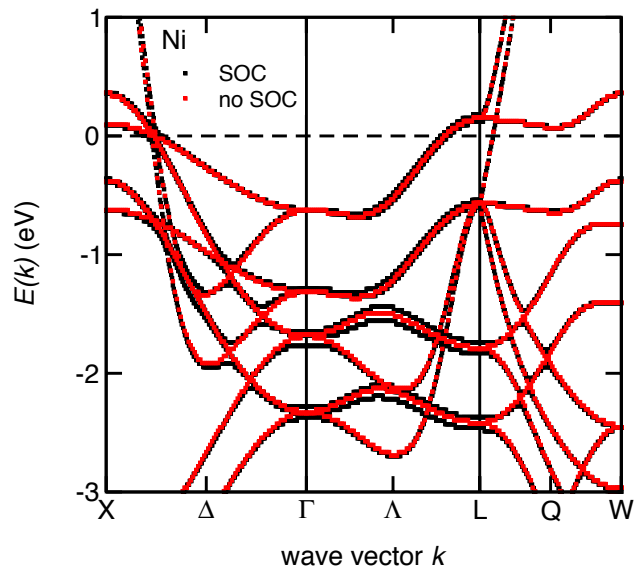

Figure 13: Band structure for fcc Ni with spin-polarization effects. Black points are with SOC and red points are without SOC. Dashed horizontal line represents the Fermi level.

Figure 14 shows the Bloch spectral function for an alloy closer to pure Ni, namely $Ni_{80}Cu_{20}$. Compared to the $Ni_{60}Cu_{40}$ shown in the bottom plot of Figure 1, the bands are less broad.

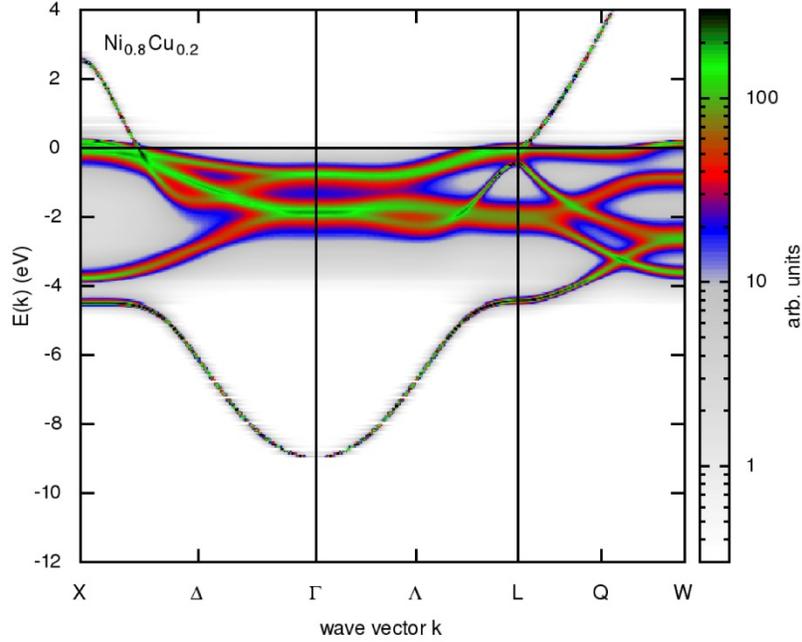

Figure 14 (color online): Bloch spectral function for $Ni_{80}Cu_{20}$ averaged over random alloy disorder configurations. The color scale is proportional to occupation probability. SOC is included and the dashed horizontal line represents the Fermi level.

## APPENDIX 2: CORRECTION FOR SHUNTING AND REMOVAL OF FARADAY CONTRIBUTION

As described in [13], obtaining accurate values of the SOT conductivities for the NM layer itself requires accounting for two effects that depend on the effective sheet impedance $Z_{eff}$ of the samples. The first of these effects is that a microwave charge current generated by SOT effects in the sample can flow in either of two paths: (1) as an image current in the CPW, where it generates a voltage across the 50 Ω impedance of the CPW and contributes to the signal measured by the VNA, or (2) through a return path within the sample itself, where it does not contribute to the measured signal. (In the limit of a thick sample with $Z_{eff} \ll 50\ \Omega$, there would be equal forward and return currents within the sample and thus no inductive coupling to the CPW.) This shunting effect, described by the schematic circuit in Figure 15, means the raw conductivities measured by our technique, $\sigma_{FL}^*$ and $\sigma_{DL}^*$, are reduced by a factor $R_s/(R_s + 50)$ from the actual SOT conductivities $\sigma_{\text{FL}}^{\text{SOT}}$ and $\sigma_{\text{DL}}^{\text{SOT}}$ (here we take $Z_{\text{eff}}$ to be the dc sheet resistance $R_s$ measured for each sample). Thus the SOT conductivities we report are

$$\sigma_{\text{FL}}^{\text{SOT}} = \sigma_{\text{FL}}^{*}\left(\frac{R_s + 50}{R_s}\right) \tag{10}$$

$$\sigma_{\text{DL}}^{\text{SOT}} = \sigma_{\text{DL}}^{*}\left(\frac{R_s + 50}{R_s}\right) \tag{11}$$

For the $R_s$ data shown in Figure 10, the shunting correction factor $(R_s + 50)/R_s$ ranges from 1.7 to 5.3.

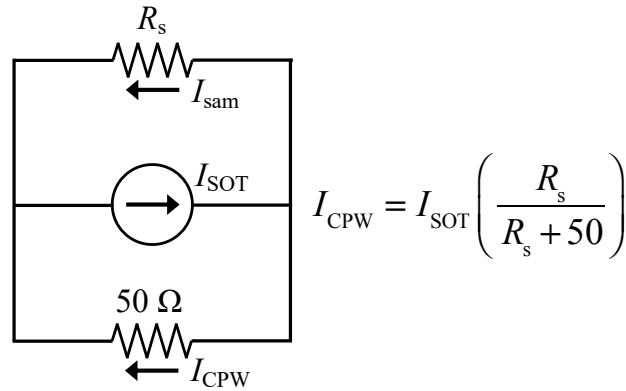

Figure 15: Schematic circuit for the shunting effect. Charge current generated by SOT in the sample can flow through the 50 Ω impedance of the CPW or through the sample itself. Only the fraction $I_{\text{CPW}}$ contributes to the signal measured by the VNA.

The second correction is needed because the fieldlike SOT and Faraday contributions to the complex inductive signal have the same phase but opposite sign (see Supplementary Information for [13]). The raw output of the inductive analysis for this signal phase, which we call the "even" conductivity $\sigma_e$ because it is even under time reversal symmetry [13], is

$$\sigma_e = \sigma_{\text{FL}}^{*} - \sigma^{\text{F}} = \sigma_{\text{FL}}^{\text{SOT}}\left(\frac{R_s}{R_s + 50}\right) - \sigma^{\text{F}}, \tag{12}$$

where we have used the shunt correction described above. Shunting does not apply to the Faraday effect because it acts as a source of electromotive force rather than a source of current (see Supplementary Information for [13]). Separating the two terms in Equation (12) relies on the fact that, for large NM thickness $d_{\text{NM}}$, we expect $\sigma_{\text{FL}}^{\text{SOT}}$ (due to interfacial effects) to be independent of $d_{\text{NM}}$ while $\sigma^{\text{F}}$ (due to currents induced by FM precession throughout the NM layer) is not. This approach is supported by the results in [14], where Cu control samples were used to subtract $\sigma^F$ at each thickness without making this assumption and the resulting $\sigma_{FL}^{SOT}$ was indeed independent of $d_{NM}$.

The Faraday term is given by [13]

$$\sigma^F = \frac{e\mu_0 m_A}{\hbar} \frac{1}{Z_{eff}}, \qquad (13)$$

where $\mu_0$ is the magnetic constant. For a given FM layer thickness, $m_A$ is fixed, while $Z_{eff}$ decreases as the NM layer thickness increases. Unlike for shunting, the actual value $Z_{eff}$ is not straightforward to determine. For an isolated NM layer of bulk resistivity $\rho_{NM}$ and thickness $d_{NM}$, we would have $Z_{eff} = \rho_{NM}/d_{NM}$, but for a metallic FM/NM bilayer this ignores the fact that some current may flow through the adjacent FM layer.

Absent an accurate model of the microwave current distribution in the bilayer, we can proceed with an empirical determination of $\sigma^F$ as follows. We measure $\sigma_e$ for a series of samples with a single FM thickness and a range of NM thickness, then plot $\sigma_e$ vs. $d_{NM}$, as shown in the top part of Figure 16 for our Py(3.5)/Ni$_{60}$Cu$_{40}$($d_{Ni60}$) samples. For the thicker films, with $d_{NM} \geq 10$ nm, the dependence is linear, as expected when the Faraday term dominates. The slope of this linear region can be described by an effective resistivity $\rho_{NM}^{eff}$ and Equations (12) and (13) then give the following expression for $\sigma_{FL}^{SOT}$:

$$\sigma_{FL}^{SOT} = \left[\frac{R_s(d_{NM})+50}{R_s(d_{NM})}\right]\left(\sigma_e + \frac{e\mu_0 m_A}{\hbar}\frac{d_{NM}}{\rho_{NM}^{eff}}\right). \qquad (14)$$

Here we have made explicit the fact that $R_s$ in the shunting correction depends on thickness, with the consequence that removing the Faraday term is not quite as simple as fitting the linear range of the top plot in Figure 16 and subtracting the slope from the entire curve. The correct removal of the Faraday term is obtained by adjusting $\rho_{NM}^{eff}$ until $\sigma_{FL}^{SOT}$ is independent of thickness for large $d_{NM}$, as shown in the bottom plot of Figure 16. In other words, the corrections for the thickness dependent Faraday term and the thickness dependent shunting must be applied together to extract $\sigma_{FL}^{SOT}$ from the raw signal.

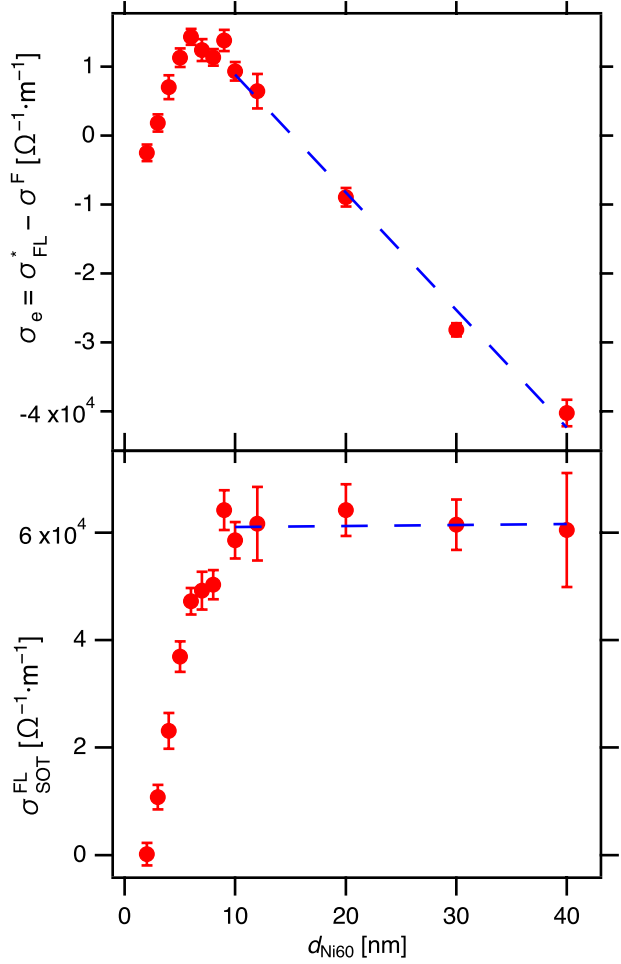

Figure 16: (top) Measured even conductivity that contains both Faraday and raw fieldlike SOT terms for Py(3.5)/Ni$_{60}$Cu$_{40}$($d_{Ni60}$) samples. (bottom) Same data after removing the Faraday term as described in the text. Dashed lines are guides to the eye highlighting the linear dependence when the Faraday term dominates $\sigma_e$ and the flat behavior of $\sigma_{FL}^{SOT}$ after removal of the Faraday term and correction for shunting.

## APPENDIX 3: COMPARISON WITH SML MODEL OF REF [29]

To facilitate a comparison with the model of Tao et al. [29], we rewrite Equation (5) in a dimensionless form,

$$\frac{G_{eff}^{tot}}{G_{\uparrow\downarrow}} = \frac{\Gamma_0 \tanh(d_{NM}/\lambda_s)}{1+\Gamma_0 \tanh(d_{NM}/\lambda_s)} + \Gamma_{SML}, \qquad (15)$$

where $\Gamma_0 \equiv G_0/G_{\uparrow\downarrow}$ with $G_0 = \sigma_0/2\lambda_s$ from Equation (4), and $\Gamma_{SML} \equiv G_{SML}/G_{\uparrow\downarrow}$. In this form, we can see that a fixed amount of spin current proportional to $\Gamma_{SML}$ is absorbed at the interface, and an additional amount proportional to $\Gamma_0/(1+\Gamma_0)$ is absorbed as the NM thickness increases from 0 to $\geq 2\lambda_s$. The dimensionless form of Equation (1) in Tao et al. is

$$\left.\frac{G_{\text{eff}}^{\text{tot}}}{G_{\uparrow\downarrow}}\right|_{\text{Tao}} = 1 - \frac{(1-\delta)^2}{1+\Gamma_0^{\text{Tao}}\tanh(d_{\text{NM}}/\lambda_s)}, \quad (16)$$

where $\delta$ is the fraction of spin current lost to SML, which is the same as our $1-\varepsilon$, and $\Gamma_0^{\text{Tao}} \equiv G_0^{\text{Tao}}/G_{\uparrow\downarrow}$ (the difference between $G_0^{\text{Tao}}$ and our $G_0$ is discussed below). Rearranging terms in Equation (16) gives a form that is directly comparable to Equation (15):

$$\left.\frac{G_{\text{eff}}^{\text{tot}}}{G_{\uparrow\downarrow}}\right|_{\text{Tao}} = \frac{\Gamma_0^{\text{Tao}}\tanh(d_{\text{NM}}/\lambda_s)}{1+\Gamma_0^{\text{Tao}}\tanh(d_{\text{NM}}/\lambda_s)} + \frac{\delta(2-\delta)}{1+\Gamma_0^{\text{Tao}}\tanh(d_{\text{NM}}/\lambda_s)} \quad (17)$$

Thus in the place of the term in our model that is independent of NM thickness, the model used in [29] has a term that decays with NM thickness,

$$\Gamma_{\text{SML}}^{\text{Tao}} = \frac{\delta(2-\delta)}{1+\Gamma_0^{\text{Tao}}\tanh(d_{\text{NM}}/\lambda_s)}. \quad (18)$$

The reason for treating SML this way is not clear to us, but it is certainly a different physical picture than a parallel conductance channel at the interface that absorbs a fixed amount of spin current.

We plot the two models in Figure 17 using the parameter values that fit the experimental data for Py/Pt in [29] and [14]. The dramatic difference between the two solid curves in the top plot is due to the fact that $\Gamma_0^{\text{Tao}}$ is nearly 70 times larger than $\Gamma_0$, which can be traced to a different prefactor for the $\tanh(d_{\text{NM}}/\lambda_s)$ term in the expression for spin absorption in the NM, Equation (4). Our model uses the conventional [18] definition $G_0 = \sigma_0/2\lambda_s$, while Tao et al. uses $G_0^{\text{Tao}} = (e^2/h)(2/3)k_F^2\lambda_m/\lambda_s$, where $k_F$ is the Fermi wavevector of Pt and $\lambda_m$ is the mean free path of Pt. Using values for Py/Pt from [29], $G_0^{\text{Tao}} = 3.70\times 10^{16}\ \Omega^{-1}\text{m}^{-2}$ and $G_0 = 2.79\times 10^{14}\ \Omega^{-1}\text{m}^{-2}$. The dashed curve in the top plot of Figure 17 shows how $\Gamma_{\text{SML}}^{\text{Tao}}$ decays with NM thickness.

To illustrate how the two models lead to different values of $\theta_{\text{SH}}$, the bottom plot of Figure 17 shows the net effect after subtracting the SML term from each model ($\Gamma_{\text{SML}}$ for the model of [14] and $\Gamma_{\text{SML}}^{\text{Tao}}$ for the model of [29]). This represents the amount of spin current that enters the NM where it can generate a SHE signal. For $d_{NM} \gg \lambda_s$, the model of Tao et al. gives a value about 3 times larger than the model used in [14]. With a much larger fraction of the spin current available to generate a SHE signal, the inferred value of $\theta_{SH}$ is necessarily smaller.

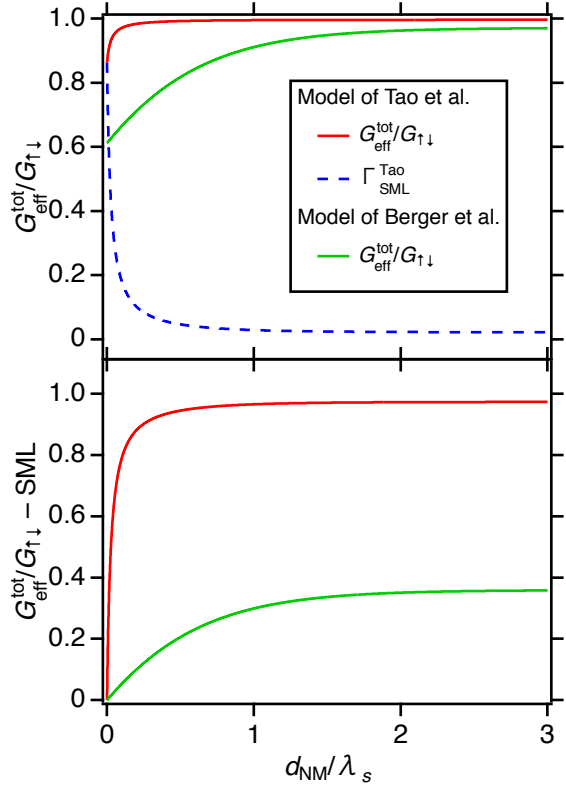

Figure 17: (top) Normalized total spin conductance vs. NM thickness for models used to fit experiments in [29] (red) and [14] (green). Dashed curve is the SML contribution for the model used in [29], given by Equation (18). (bottom) Subtracting the SML contribution gives the net spin current available for spin-to-charge conversion in the NM layer. Parameters taken from [29]: $\Gamma_0^{Tao} = 38$ and $\delta = 0.63$. Parameters taken from [14]: $\Gamma_0 = 0.56$ and $\Gamma_{SML} = 0.61$.

---